\documentclass[12pt]{JHEP3} 

\usepackage{epsfig,multicol,bbm}
\usepackage{graphicx}
\usepackage{amssymb}
\usepackage{slashed}

\title{Higgs Signal for $h \to aa$ at Hadron Colliders}

\author{Marcela Carena\\
	Fermi National Accelerator Laboratory,
	Batavia, IL 60510, USA \\
	E-mail: \email{carena@fnal.gov}}
\author{Tao Han~
		and Gui-Yu Huang\\
	Department of Physics, University of Wisconsin, 
	Madison, WI 53706, USA \\
	E-mail: \email{than@hep.wisc.edu}, \email{ghuang@hep.wisc.edu}}
\author{Carlos E.M.~Wagner\\
	HEP Division, Argonne National Laboratory,Argonne, IL 60439, USA\\
	Enrico Fermi Institute and Kavli Institute for Cosmological Physics, \\
        Physics Department, University of Chicago, Chicago, IL 60637, USA\\
	E-mail: \email{cwagner@hep.anl.gov}}

\preprint{
FERMILAB-PUB-07-652-T \\
MADPH--07--1497 \\
ANL-HEP-PR-07-106\\
EFI-07-39
}

\abstract{We assess the prospect of observing a neutral Higgs boson
at hadron colliders in its decay to two spin-zero states, $a$,  
for a Higgs mass 
of $90-130$ GeV, when produced in association with a $W$ or $Z$ boson. 
Such a decay is allowed in extensions
of the MSSM with CP-violating interactions and in the NMSSM, 
and can dominate Higgs boson final states, thereby evading the 
LEP constraints on standard Higgs boson production.
The light spin-zero state decays primarily via
$a\rightarrow b\bar{b}$ and $\tau^+\tau^-$, so this signal channel
retains features distinct from the main backgrounds.
Our study shows that at the Tevatron, there may be potential to observe a few
events in the $b\bar{b}\tau^+\tau^-$ or $b\bar{b}b\bar{b}$ channels
with relatively small background, although this observation would be statistically 
limited.  At the LHC, the background problem is more severe, but with
cross sections and integrated luminosities orders of magnitude larger
than at the Tevatron, the observation of a Higgs boson in this decay mode
would be possible.  The channel $h\rightarrow aa\rightarrow b\bar{b}b\bar{b}$
would provide a large statistical significance, with a signal-to-background 
ratio on the order of $1:2$. In these searches, the main challenge would be 
to retain the adequate tagging efficiency of $b$'s and $\tau$'s in the 
low $p_T$ region.}


\begin{document}

\def\mh{m_h^{}}
\def\gev{{\rm GeV}}
\def\tev{\rm TeV}
\def\fbi{\rm fb^{-1}}
\def\abi{\rm ab^{-1}}
\def\ee{e^+e^-}
\def\nn{\nu\bar\nu}
\def\ttb{t\bar t}
\def\tth{t\bar t h}
\def\lsim{\mathrel{\raise.3ex\hbox{$<$\kern-.75em\lower1ex\hbox{$\sim$}}}}
\def\gsim{\mathrel{\raise.3ex\hbox{$>$\kern-.75em\lower1ex\hbox{$\sim$}}}}
\def\simlt{\stackrel{<}{{}_\sim}}
\def\simgt{\stackrel{>}{{}_\sim}}

\newcommand{\beq}{\begin{equation}}
\newcommand{\eeq}{\end{equation}}
\newcommand{\bea}{\begin{eqnarray}}
\newcommand{\eea}{\end{eqnarray}}

\section{Introduction}

The elucidation of the mechanism leading to the origin of mass
of all observed elementary particles is one of the main goals in
high energy physics. The simple Standard Model (SM) picture, based
on the spontaneous breakdown of the electroweak symmetry by the
vacuum expectation value (vev) of an elementary Higgs field, seems to
lead to a picture that is consistent with all experimental observables,
provided the Higgs boson mass is smaller than about
$250\,\gev$. Moreover, the best fit to the precision electroweak observables 
measured at the LEP, SLC and Tevatron experiments lead to 
values of the Higgs mass of the order of or
smaller than the present bound coming from direct searches at LEP,
$m_{H_{\rm SM}} \simgt 114\,\gev$~\cite{Barate:2003sz}.

In spite of the extraordinary good agreement of the experimental
observations with the SM predictions, there are many
theoretical motivations to go beyond the SM description.
Several extensions of the SM exist in the literature,
and in most of them the Higgs sector is extended to a more complicated
structure, often including at least two Higgs doublets. The requirement
of preserving the good agreement with experimental data can be
easily fulfilled in extensions, like supersymmetry, in which the
effect of the additional particles on the precision electroweak observables
rapidly vanish with increasing values of the new particle masses.
An extension of  the Higgs sector will generally require a revision of 
the direct and indirect limits on the Higgs mass. In particular, the
direct search for Higgs bosons may be affected by additional decay
modes that are beyond the ones analyzed by the LEP collaborations.

As an example, let us consider the minimal supersymmetric extension of
the SM (MSSM). In the MSSM,
there is an additional Higgs doublet, 
leading, in the absence of CP-violation in the Higgs sector, 
to two CP-even and one CP-odd Higgs 
boson states.  At large values of $\tan\beta$, the ratio of 
 of the two Higgs doublets vev's,
one of the CP-even Higgs bosons acquires SM properties, while the
second Higgs boson may be produced in association with the CP-odd
Higgs boson state. In addition, the masses of the non-standard CP-even
Higgs and the CP-odd Higgs are close to each other. Under these conditions,
the mass bound on the SM-like CP-even Higgs is similar to the SM one, while
the CP-odd and the second CP-even Higgs boson mass bound reads
$m_h\gsim 90\,\gev$~\cite{Heister:2001kr}. 

In this paper, we will depart from these simple assumptions, by 
breaking the mass relations that appear in the simplest supersymmetric
models, and studying the consequences of such modifications of the
parameters of the theory. 
Indeed, while it has been a common belief that the Higgs boson 
will be eventually discovered at the upcoming LHC experiments, 
one would like to fully utilize the potential 
to search for the Higgs bosons at the Tevatron in these non-conventional
scenarios as well. Non-standard mass relations are already present 
in  extensions of the  MSSM  including 
an additional singlet (NMSSM~\cite{Dermisek:2005ar,Han:2004yd} and other
extensions~\cite{Cvetic:1997ky,Panagiotakopoulos:1999ah,Panagiotakopoulos:2000wp,Dobrescu:2000yn,Erler:2002pr,Barger:2006rd}), 
or when explicit CP-violation exist in the Higgs sector~\cite{Carena:2002bb}.
In these cases, the SM-like Higgs ($h$) may  dominantly 
decay into a pair of lighter Higgs ($a$), 
an admixture of CP even and odd states with a dominant CP odd component.
(The precise fraction of the CP even or odd component is not crucial in the present study.)
Therefore it is possible that the Higgs escaped detection at the 
LEP experiments
by avoiding the usual decay modes such as 
$h\to b\bar{b},\ \tau^+\tau^-,\ WW^*$ and $ZZ^*$, 
and the lower limit on Higgs mass should be 
re-evaluated. The LEP collaborations have already analyze the possible
constraints on Higgs boson production arising from this new decay 
mode~\cite{Abdallah:2004wy,Schael:2006cr}. We shall use the results
of these analyses as a starting point for our study. We are interested 
in analyzing the
sensitivity of the Tevatron and the LHC experiments in the search for  
a light, SM-like  Higgs boson with such an exotic decay mode.

We consider the case where the SM-like Higgs boson decays into 
a pair of spin-zero states, $h\to aa$, which in turn cascade into a heavy 
fermion pair $a\to b\bar{b}$ or $a\to \tau^+\tau^-$. 
These Higgs-to-Higgs decay modes have been studied extensively in 
NMSSM~\cite{Ellwanger:2003jt,Ellwanger:2005uu,Chang:2005ht}
at the LHC, together with even more complicated cascades.
Most of these studies indeed take advantage of the dominant 
production modes of the Higgs boson at hadron colliders, i.e. gluon 
fusion and weak boson fusion, but encounter large SM backgrounds.
We therefore consider the Higgs signal produced in association 
with a $W$ or $Z$ boson, 
where the leptonic decays of the weak bosons will
provide a clean trigger, and will significantly reduce the background
as well.~\footnote{An initial analysis by us at the Tevatron was reported 
earlier in Ref.~\cite{Aglietti:2006ne}. 
While this current work was in process, another similar analysis for the 
$4b$ channel at the LHC appeared \cite{Cheung:2007sva}. 
For the overlap with that work at the LHC,
our studies included more background analyses, more realistic
$b$-tagging effects, and a broader parameter scan.}

The significance of associated production was also stressed in 
Ref.~\cite{Moretti:2006hq} for $h\to aa$ decays in NMSSM at the LHC. 
For similar parameter choices, our results lead to a comparable or slightly
better reach for Higgs boson searches to the ones obtained in earlier
studies~\cite{Ellwanger:2003jt,Ellwanger:2005uu,Chang:2005ht} on 
gluon fusion and weak boson fusion productions at the LHC,
while showing more discovery potential at the Tevatron than in 
previous studies.

\section{Signal Processes and Parameter Choices}

\subsection{Signal Processes}

For the two spin-zero states, $a$'s,
the combinations of decay products we can search for are 
$4b$, $2b2\tau$ and $4\tau$. The $4\tau$ mode is usually 
suppressed by the branching fractions, unless $m_a$ 
is below the $b\bar{b}$ decay threshold~\cite{Graham:2006tr}.
We thus concentrate on the two channels 
$2b 2\tau$ and $4b$ next. 
The signal events being searched are
\bea
\label{wh}
Wh\to & l\nu_l, aa &\to \left\{ \begin{array}{l}
l\nu_l, b\bar{b}, b\bar{b} \\
l\nu_l, b\bar{b}, \tau^+\tau^-
\end{array} \right. \\
Zh\to & l^+l^-, aa &\to \left\{ \begin{array}{l}
l^+l^-, b\bar{b}, b\bar{b} \\
l^+l^-, b\bar{b}, \tau^+\tau^-,
\end{array} \right. 
\label{zh}
\eea
with $l=e,\mu$.
The channel $Z\to\nu\nu$ decays into neutrino pairs can also be considered,
while the triggering could be large missing energy, plus $\tau$'s or $b$'s.  

\subsection {Parameter Choices}
We would like to perform a relatively model-independent search for 
the signal, therefore the Higgs masses, branching fractions and
couplings to the weak bosons are employed as input parameters.
Direct searches for a Higgs boson with SM-like couplings to the
gauge bosons, in a model and decay mode-independent way, 
leads to a lower bound on $m_h$ of about 
$82\,\gev$~\cite{Abbiendi:2002qp} with full SM coupling to $Z$. 
On the other hand, the proposed search
is expected to become inefficient for $m_h>130\,\gev$, since
the standard decays into the $WW^*$ and $ZZ^*$ channels
are still expected to be dominant.  Therefore, 
the optimal setting to detect the Higgs decaying into an $aa$ pair is to 
have the mass $m_h$ within the range of $90-130\,\gev$. 
The choice for $m_a$ can be more flexible.  
As long as $m_h>2m_a$ and $m_a>2m_b$ to kinematically allow the decays 
$h \to aa$ and $a \to b\bar{b}$, our methods
are rather insensitive to the mass choices.

In a generic model, the $Wh/Zh$ production rate differs from that in the SM.
The change can be characterized by a prefactor $\kappa^2_{hWW}$
($\kappa^2_{hZZ}$), where $\kappa_{hVV}$ is the coupling strength of Higgs to
vector boson $V$ relative to that in the SM.  The production cross section can
thus be written in terms of the SM result with an overall factor to account for
the modification of the coupling
\bea
\sigma(Vh) = \kappa^2_{hVV} \sigma^{SM}(Vh) .
\eea
We are interested in the range of
$\kappa^2 \sim 0.5-1.0$, so that this Higgs contributes to the
electroweak symmetry breaking and consequently the 
associated productions are still sizable.

In order for the $h\to aa$ decay to be dominant and thus escape the LEP bound, 
$BR(h\to aa)$
is required to be close to unity. For instance, in the NMSSM, 
$BR(h\to aa)>0.9$ turns out to be very
general in terms of the naturalness of $c$
in the trilinear coupling term $(cv/2)haa$~\cite{Dobrescu:2000jt}.
Moreover, if the down quark and lepton coupling to the Higgs
is proportional to their masses, then 
$BR(a\to b\bar{b})$ and
$BR(a\to \tau^+\tau^-)$ are set to be 0.92 and 0.08, respectively.
In general, however, the relations between the coupling and the masses
may be modified by radiative corrections, which can lead to a large
increase of the $BR(h \to \tau\tau)$~\cite{Carena:1998gk}.
The representative values and the ranges of the parameters
are summarized in Table \ref{param}, all allowed by constraints from
LEP~\cite{Abdallah:2004wy,Schael:2006cr}, except for the region near 
$m_h \sim 90\ \gev$ when both $a$'s are assumed to decay into two
bottom quarks.
Parametric consistency with the LEP results is also discussed 
in detail~\cite{Dermisek:2005gg} within the NMSSM framework.

\TABLE{
\begin{centering}
    \begin{tabular}[t]{ l || l ||  c|c}
      \hline
       &   & representative & considered  \\
       & parameters  & value & range  \\
      \hline \hline 
            & $m_h$ & ~120~ & ~90$-$130~   \\
      \raisebox{1.5ex}[0pt]{masses}
            & $m_a$ & ~30~ & ~20$-$60~   \\
      \hline
	coupling
            & $\kappa^2_{hVV}$ &  ~0.7~  & ~0.5$-$1.0~  \\
      \hline
            & $BR(h\to aa)$ &  ~0.85~  & ~0.8$-$1.0~  \\
      \raisebox{1.5ex}[0pt]{branching}
            & $BR(a\to b\bar{b})$ &  ~0.92~ & ~ 0.95$-$0.50 ~ \\
      \raisebox{1.5ex}[0pt]{fractions}
            & $BR(a\to \tau^+\tau^-)$
 &  ~0.08~ & ~ 0.05$-$0.50 ~ \\
      \hline
      \hline
	$2b2\tau$ channel & $C^2_{2b2\tau}$ &  ~0.088~  & ~0.038$-$0.50~  \\
	$4b$ channel & $C^2_{4b}$ &  ~0.50~  & ~0.10$-$0.90~  \\
      \hline       \hline
    \end{tabular}
    \caption{Parameter choices for $h\to aa$ decays. 
         The $C^2$ factor is defined in the next section.
    \label{param}}
\end{centering}
}

\section{$h \to aa$ at the Tevatron}

\subsection{The $2b 2\tau$ Channel}

Including the decay branching fractions for
$aa\to b\bar{b}  \tau^+ \tau^-$, 
we obtain the  cross section as
\begin{eqnarray}
   \sigma_{2b2\tau} =  \sigma(Vh) \ BR(V)\  
	2BR(h\to aa) BR(a\to b\bar{b}) BR(a\to\tau^+\tau^-) .
    \label{aa2b2tauxsec}
\end{eqnarray}
where $BR(V)=0.213\ (0.067)$ is
the leptonic branching fraction of $W\ (Z)$
decay into $l=e,\mu$.

The overall factor modifying the SM result 
in Eq.~(\ref{aa2b2tauxsec}), 
\beq
C^2_{2b2\tau} \equiv 2\kappa^2_{hVV}BR(h\to aa) BR(a\to b\bar{b}) BR(a\to\tau^+\tau^-),
\eeq 
corresponds to the process-dependent $C^2$ factor defined in 
the DELPHI search~\cite{Abdallah:2004wy}, and the $S_{95}$ factor in the comprehensive
LEP analysis~\cite{Schael:2006cr}. 
Our parameter choice (range), as listed in Table~\ref{param}, 
is equivalent to 
a $C^2_{2b2\tau}$  of 0.088 (0.038$-$0.50), 
consistent with the bounds for a large range of our $m_h,m_a$ choices
set forth in Refs.~\cite{Abdallah:2004wy,Schael:2006cr}.
A value of 0.088 for $C^2_{2b2\tau}$ is assumed 
for all numerical evaluations from here on, unless explicitly noted otherwise.

\subsection{Signal Event Rate for the $2b2\tau$ Channel}
The associated production of $p\bar{p}\to Wh$ 
usually features a larger cross section than that of $Zh$,
and the leptonic branching fraction of $W$ is 
about 3 times larger than $Z$'s.
For illustration purposes, we choose to present our detailed studies for 
the $Wh$ channel henceforth, although we will include the $Zh$ channel
in our results. 
\FIGURE[t]{
    \includegraphics[scale=0.34]{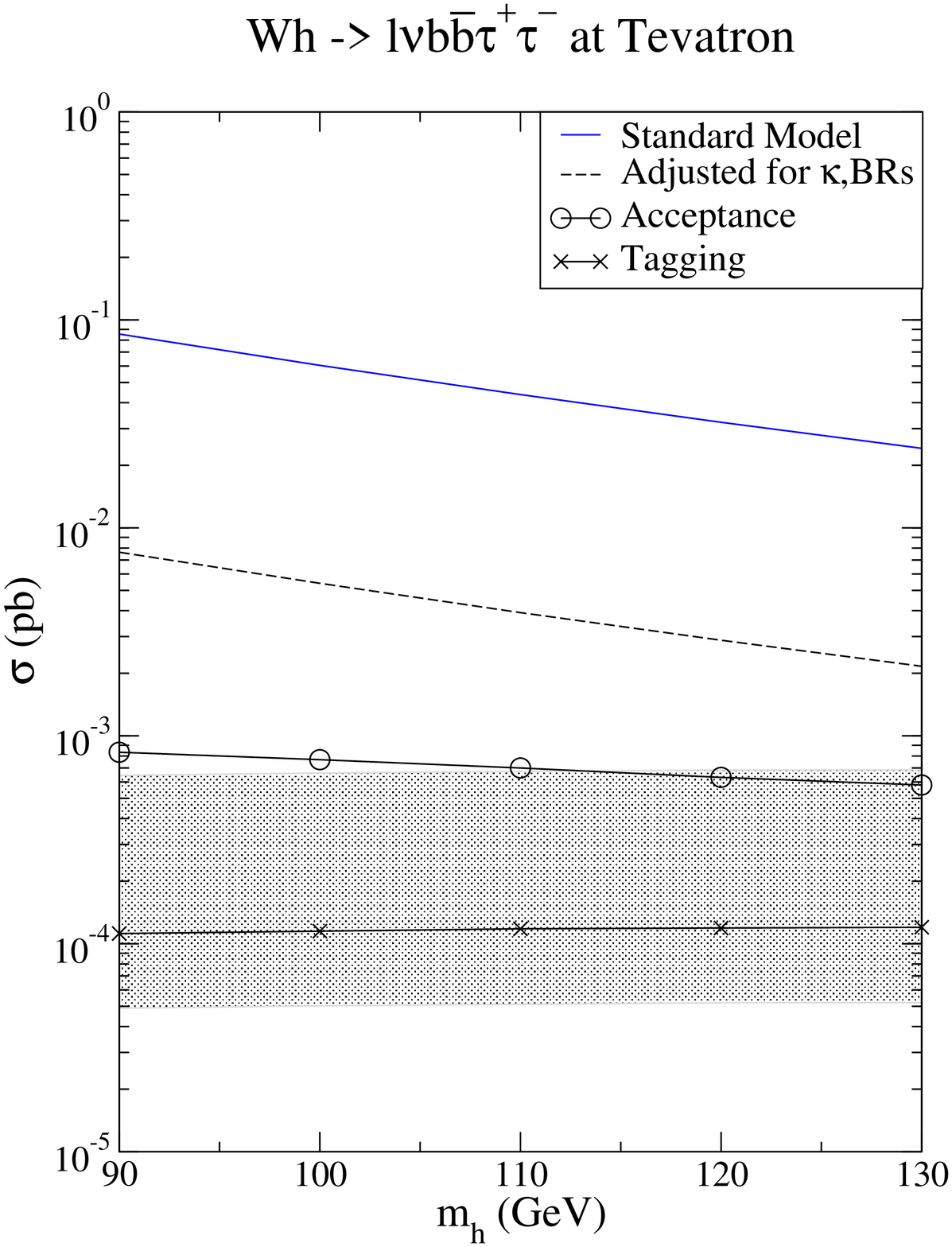}
    \includegraphics[scale=0.34]{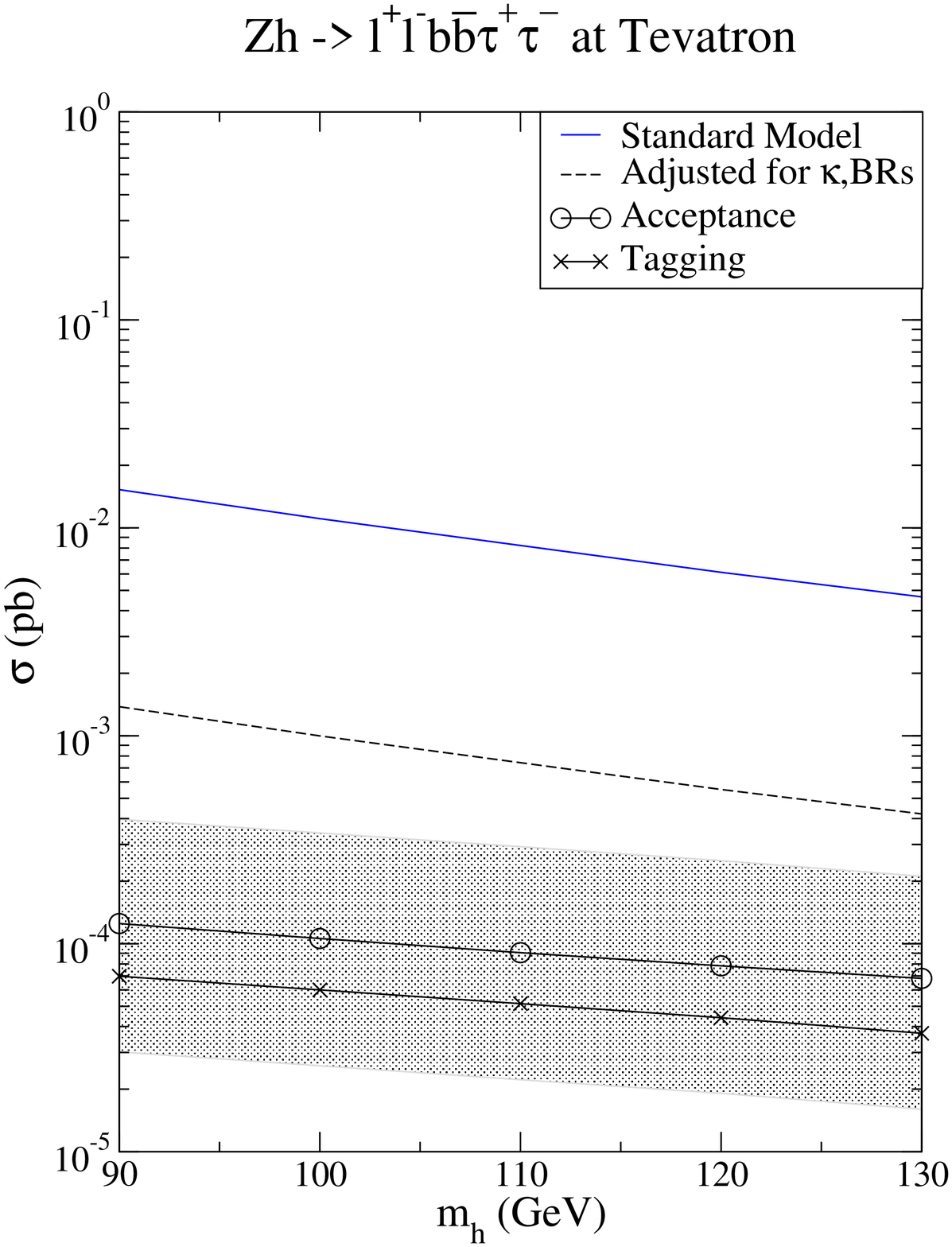}
    \caption {Cross sections of Higgs signal at the Tevatron
	in the $2b2\tau$ 
	channel produced by Higgs-strahlung with a leptonically 
	decaying $W$ (left) or $Z$ (right). 
The four lines from the top to the bottom correspond to, respectively, 
the SM cross section (solid line); adjusted for 
$C^2_{2b2\tau}=0.088$ and $m_a=30\,\gev$ (dotted line); 
including the  acceptance cuts Eqs.~(\ref{eq:TeVcut1})$-$(\ref{eq:TeVcut2})
(line with circles); 
and further including tagging efficiencies Eq.~(\ref{taggingTeV}) (line with crosses).
 The shaded bands  correspond to variations  of the final results (line with crosses)
for values of $C^2$ 
within the range considered in Table~\ref{param} and 
taking into account the LEP constraints \cite{Schael:2006cr}. }
    \label{xsec}
}

Total cross sections for $p\bar{p}\to Wh$ and $p\bar{p}\to Zh$ at 
hadron colliders have been calculated 
with QCD and electroweak corrections 
included~\cite{Han:1991ia,Ciccolini:2003jy,Hahn:2006my}
in the SM. Hence we get
\beq
 \sigma^{SM}(Wh) \ BR(W\to l\nu_l) \sim 85\ (24) \ {\rm fb}
\eeq
 at $\sqrt s=1.96$ TeV for $m_h=90\ (130)\,\gev$.

Including the branching fractions and couplings,
the cross section of the signal in Eq.~(\ref{aa2b2tauxsec}) is
\beq
 \sigma_{2b2\tau} \sim 7.5\ (2.1) \ {\rm fb \qquad for \quad} C^2 = 0.088
\eeq
as illustrated in Fig.~\ref{xsec}.
The solid curve on top represents the total cross section for $Vh$ production, 
with $V$ decaying leptonically, but without any cuts.
The dashed curve represents the cross section after 
adjusting for the couplings and branching fractions, 
as in Eq.~(\ref{aa2b2tauxsec}). 
Cross sections for $Zh$ are also plotted for completeness.

\subsection{Background and Cuts for the $2b2\tau$ Channel}

The main advantage for considering the process $Wh$ is the possible background
suppression due to the clean final state
from the $W$ leptonic decay: an isolated charged lepton ($l=e,\ \mu$) plus
large missing transverse energy. 
We thus require the following initial acceptance cuts at the Tevatron
~\cite{Abazov:2006hn}
\beq
p_T(l)>15\,{\gev} , \quad 
|\eta(l)|<2.0, \quad 
\slashed{E}_T>15\  {\gev }.
\label{eq:TeVcut1}
\eeq
The events have yet to further pass the acceptance cuts, or to have the taus and $b$'s tagged. 
Both help suppress the SM backgrounds,  while
bringing significant reductions to the event rate as well. 
Our challenges are to retain as many signal events as possible, and to control
the backgrounds from various sources.
Throughout the paper, we adopt the Monte Carlo program 
MadEvent~\cite{Maltoni:2002qb} for our background simulations
at the parton level.

\paragraph{$b$ and $\tau$ Tagging}

We wish to identify events with 5 particles plus missing energy 
in the final states:  $b\bar{b}\tau^+\tau^- l\nu_l$. 
With neutrinos in the decay products, tau momenta cannot be fully reconstructed.
Therefore we cannot reconstruct the invariant masses $m(2\tau)$ or 
$m_h\sim m(2b2\tau)$. 
Instead, the signal should appear as a peak
in the $m(2b)$ plot, around the value of $m_a$.

For the jets and other soft leptons in the events, 
the following basic  cuts are employed to mimic 
the CDF~\cite{Lukens:2003aq} detector acceptance, 
for jets~\cite{Aaltonen:2007dm}:
\beq
p_T(j)>10\,\gev , \quad
|\eta(j)|<3.0 ,
\label{accept-cut}
\eeq
and for $\tau$-candidates~\cite{Abulencia:2007kq}:
\beq 
p_T>10,8,5\,\gev\ {\rm for }\ \tau_h, \tau_e, \tau_{\mu}, \quad
|\eta|<1.5.
\label{tau-cut}
\eeq
where 
$\tau_e$, $\tau_\mu$ and $\tau_h$ stand for the visible decay products of
$\tau\to e \nu_e \nu_\tau$,
$\tau\to \mu \nu_\mu \nu_\tau$, and
$\tau\to \rm{ hadrons }+ \nu_\tau$,
respectively, and an isolation cut
\beq
\Delta R > 0.4
\label{eq:TeVcut2}
\eeq
 between leptons, $\tau$'s and $b$-jets. 
After the acceptance cuts,
$10-25$\% of the signal events survive, and the cross section becomes 
$0.85 \ (0.57)\ $fb for $m_h=90\ (130)\, \gev$ 
with the given set of input parameters ($C^2\sim 0.088$). 
The cross sections passing acceptance
are plotted in Fig.~\ref{xsec} versus the Higgs mass, represented by the circled curve. 
At this level, the cross section is below 1~fb.

The $b$- and hadronic $\tau$-tagging efficiencies and the kinematics
are taken to be
\bea
\nonumber
\epsilon_b = 50\%  &{\rm for}\ E_T^{jet}>15\,\gev \ &{\rm and} 
\ |\eta_{jet}|<1.0\ , \\
\epsilon_\tau= 40\% &{\rm for}\ E_{vis}>20\,\gev \ &{\rm and} \ |\eta|<1.5\ . 
\label{taggingTeV}
\eea
Outside these kinematical regions, 
the tagging efficiencies drop off sharply~\cite{Acosta:2005ij,Jeans:2005ew}.
We decide to tag one $b$ and one tau.
The energies for a jet and a lepton are smeared according a Gaussian distribution.
The energy resolutions  are taken to be
\bea
{\Delta E_j\over E_j} = {75\% \over \sqrt{E_j}} \oplus 5\% ,\quad
{\Delta E_l\over E_l} = {15\%\over \sqrt{E_l}} \oplus 1\% .
\eea
The missing energy is reconstructed according to the smeared observed particles.
No further detector effects are included \cite{Abulencia:2005ix}.

\paragraph{Irreducible Background}

The dominant source of the irreducible background, 
with the same final state as the signal,
\beq
 W\ Z^*/\gamma^*(\to \tau^+\tau^-)\ b\bar{b},
 \label{irre}
\eeq
has the $b\bar{b}$ pair from a virtual gluon splitting, 
the $\tau^+\tau^-$ pair from an intermediate $Z^*/ \gamma^*$ and
the charged lepton plus missing energy from a $W$ boson. 
Our simulations show that the largest contribution come from events with
the $Z^*$ almost on-shell, while the 
 $\tau^+\tau^-$ pair from a virtual photon 
can be more easily confused with the signal.
After applying the acceptance cuts, 
the irreducible background is estimated 
to be around 0.01 fb,
which is very small compared to the signal size.
It is essentially absent given the luminosity expected at the Tevatron.

\paragraph{Reducible Background}

Reducible background arise from jets mis-identified as $b$'s, or
as hadronically decaying taus. The mistag rate from a light quark is taken to be 
$0.5-1.0$\% for tau and $0.5$\%  for $b$, respectively \cite{Acosta:2005ij,Jeans:2005ew}. 
A charm quark has higher mistag probability to fake a $b$ quark, that we take to be 
$10\%$~\cite{Acosta:2004hw}. 
In addition, the experiments cannot distinguish directly produced electrons (muons) from
leptonically decaying taus. Thus the reducible backgrounds considered
in our study are listed below.
\begin{itemize}
\item 
The background due to misidentified bottom comes from the process
$2\tau 2j\  l + \slashed{E}_T$, 
which has a cross section of 5 fb. Considering the mistag rate and 
the additional cuts, it contributes 0.02 fb to the background events.
\item 
The background due to misidentified $\tau$
differs for different decay modes of $\tau$'s: 
  \begin{itemize}
    \item
For
$\tau_l \tau_h 2b l \slashed{E}_T$ ($2l2b \tau_h \slashed{E}_T$),
it comes from $2\tau 2b j$ with
$\slashed{E}_T$ from the leptonic decays of both taus. 
The contribution is estimated at 
0.003 fb.

    \item
For
$\tau_h \tau_h 2b l \slashed{E}_T$, the background
comes from $2j 2b l \slashed{E}_T$ and is estimated at $30$ fb. 
It is then reduced by the tau-mistag rate, the b-tagging rate, 
and their associated cuts. In the continuum distribution of 
$m(2b)$, it is below the level of the resonant signal. 
Within the mass window of $10\,\gev < m(2b) <70\,\gev$, this background
accumulates to $0.04-0.09$ fb, depending on the $\tau$ mistag rate considered.
  \end{itemize}

\item
The backgrounds from both a mistagged tau and a mistagged $b$ 
mostly come from the $4j l \slashed{E}_T$ events, 
which has a cross section of about 16 pb.
After the cuts and folding in the mistag rates, 
this contributes $0.03-0.05$ fb of background 
in $10\,\gev < m(2b) <70\,\gev$, depending on the $\tau$ mistag rate considered.
\end{itemize}

The two bottom-quarks in the final state coming from the Higgs
boson decays should have an invariant mass equal $m_a$. If 
enough data were available, one would be able to
observe an excess of events in the $m(2b)$ mass
distribution. However, this procedure is heavily limited by
statistics. For instance,
with a window cut of $m_a \pm 10\, \gev$ on $m(2b)$,
the reducible background can be a factor of 3 to 5
smaller than the signal,
but unfortunately, the cuts and the tagging efficiencies 
together reduce the signal greatly to about 0.11 fb for $Wh$ and
$0.05-0.07$~fb for $Zh$, with $C^2 \sim 0.088$ as shown in Fig.~\ref{xsec}
by the crossed curve. 
The shaded band represents the range of parameters
allowed by our choice of $C^2 \sim 0.038-0.50$, consistent with the LEP 
constraints.
With an optimistic value of $C^2 \sim 0.50$, 
the cross section is 0.68 fb,
and we would expect to see about a couple of signal events
with an integrated luminosity of a few $\fbi$. 

To illustrate a most optimistic situation in terms of kinematical
 considerations, we explore the 
optimization between $m_a$ and $m_h$ to obtain the largest signal rate.
The signal loss is mainly due to the softness of the $b$ and $\tau$'s, 
therefore most events are rejected from the lower $p_T$ threshold. 
Increasing $m_a$ would stretch the $p_T$ distributions to the higher
$p_T$ end.
To achieve this without significantly affecting the decay phase space
of $h$, we set 
\beq
m_a=(m_h-10\,\gev )/2,
\label{mamh}
\eeq
which resulted in more than doubling the signal rate with respect
to the $m_a = 30\,\gev$ case as our default presentation throughout.
In this case the signal cross section is $\sim 0.28\ {\rm fb}$ for 
$C^2=0.088$,  and $\sim 1.6\ {\rm fb}$ for $C^2=0.50$, which 
is still challenging for observation 
with the Tevatron's projected luminosity.

\subsection{The $4b$ Channel}

\FIGURE[t]{
    \includegraphics[scale=0.34]{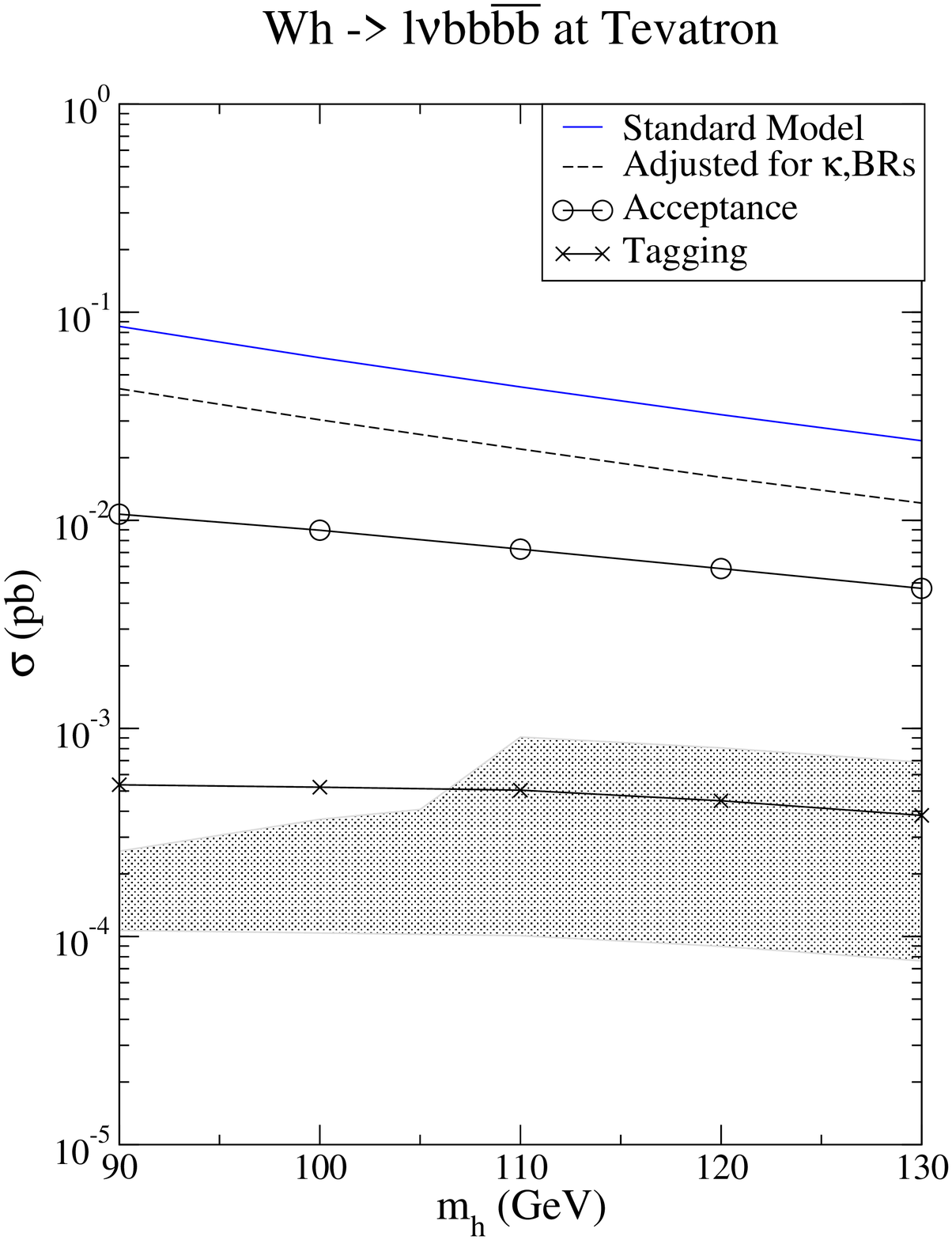}
    \includegraphics[scale=0.34]{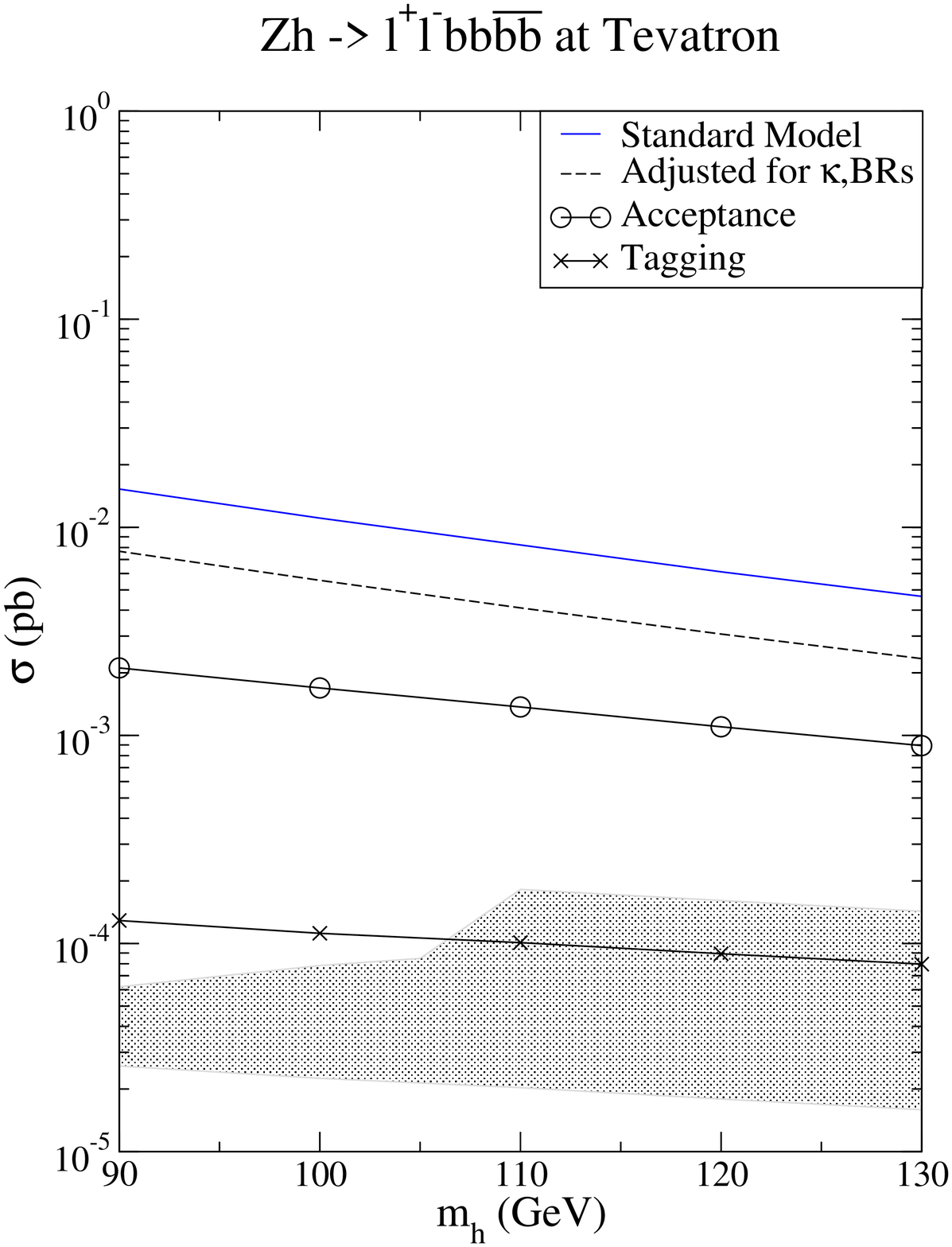}
    \caption {
    Cross sections of Higgs signal at the Tevatron
	in the $4b$ 
	channel produced by Higgs-strahlung with a leptonically 
	decaying $W$ (left) or $Z$ (right).
The four curves from top to bottom correspond to, respectively, 
the SM cross section (solid line); adjusted for 
$C^2_{4b}=0.50$ and $m_a=30\,\gev$ (dotted line); 
including the  acceptance cuts 
Eqs.~(\ref{eq:TeVcut1}), (\ref{accept-cut}) and (\ref{eq:TeVcut2})
(line with circles); 
and further including tagging efficiencies Eq.~(\ref{taggingTeV}) (line with crosses).
 The shaded bands  correspond to variations  of the final results (line with crosses)
for values of $C^2$ 
within the range considered in Table~\ref{param} and 
taking into account the LEP constraints \cite{Schael:2006cr}. }
    \label{xsec4b}
}

As we mentioned earlier, the light spin-zero state $a$ mostly decays
into $b\bar{b}$ ($50-95\%$) or $\tau^+\tau^-$ ($5-50\%$). 
The $\tau^+\tau^-$ channel can be dominant when $a$ is very light,
{\it i.e.} $m_a \simlt 2m_b$. 
However it would be difficult to observe Higgs in the $4\tau$ mode, 
first because of the increasing
background near the lower end of $m(\tau\tau)$, 
and because of the difficulty in resolving highly-collimated tau pairs.
These scenarios involving a very light $a$ are among the difficult 
ones for NMSSM Higgs discovery discussed in Ref.~\cite{Ellwanger:2005uu}.
A relevant $4\tau$ study at the Tevatron under such scenario can be found 
in Ref.~\cite{Graham:2006tr}.
A very light $a$ could also be probed through Upsilon or 
even $J/\Psi$ decays~\cite{Dermisek:2006py}.
For the $m_a >2m_b$ case, we will next look for
the Higgs in the $4b$ channel.

Similar to the $2b2\tau$ mode, the $4b$ cross section is
\begin{eqnarray}
   \sigma_{4b} =  \sigma(Vh) \ BR(V)\  
	BR(h\to aa) BR(a\to b\bar{b})^2 ,
    \label{aa4bxsec}
\end{eqnarray}
from which we extract the $C^2$ factor
\beq
C^2_{4b} \equiv \kappa^2_{hVV}BR(h\to aa) BR(a\to b\bar{b})^2.
\eeq 

The $4b$ mode is usually enhanced by the large branching fractions 
of the decay of $a$ into bottom quarks.
The ratio $C^2_{4b}/C^2_{2b2\tau}$ 
ranges in $9.5-0.5$ for $BR(a\to\tau\tau)\sim 0.05 - 0.50$. 
The value of
$C^2_{4b}$ itself does not vary greatly with the branching fractions
that are obtained  within our choice of parameters, Table~\ref{param}.
Despite larger background for this mode than for the $2b2\tau$ mode,
the enhanced rate suggests this to be a more viable mode.

The parameter choices for the $4b$ channel are
also given in Table.~\ref{param}.
Running parallel to the $2b2\tau$ channel,
we plot the cross sections in Fig.~\ref{xsec4b}.
The shaded bands show the  LEP constraints disfavoring
 the lower end of the $m_h$ range.
We find the signal rate after acceptance cuts to be $10.7-4.7$ fb 
(the circled curve)
for $m_h=90-130\,\gev$ with $C^2\sim 0.5$. 
After tagging three bottom jets (for reasons explained below) 
and imposing appropriate additional cuts,
the cross section becomes $0.54-0.38$ fb (the crossed curve)
for $m_h=90-130\,\gev$.

We again adopt the basic acceptance cuts and the $b$-tagging 
requirements as in the previous section. 
The background for this mode arises from $4bl\slashed{E}_T$,
$3bjl\slashed{E}_T$,$2b2jl\slashed{E}_T$, $b3jl\slashed{E}_T$ and
$4jl\slashed{E}_T$ events.
For the four $b$'s in our signal, tagging two will not be sufficient,
as background from $2b2jl\slashed{E}_T$ events can fake the signal 
without any mistagging involved.
Therefore we demand that at least three bottom jets be tagged.

The irreducible background  $4bl\slashed{E}_T$, 
though much larger than that in the $2b2\tau$ mode,
is still manageable. 
We find the cross section to be 0.23 fb after basic acceptance cuts.
Like the signal events, 
it suffers similar reductions from tagging and further cuts,
which brings it down to 0.02 fb. With tagging for 3$b$'s, 
the $3bjl\slashed{E}_T$ events cannot be effectively distinguished from the signal 
either. They contribute about 0.003 fb to the background.

The reducible backgrounds from $2b2jl\slashed{E}_T$ and
$4jl\slashed{E}_T$ events have the same sources as that in 
the $2b2\tau$ mode and 
the mistag rates of a light jet to $ b$ and to $ \tau$ are comparable.
The tagging on the 3$^{rd}$ $b$ brings this background down
significantly. In total, they contribute about 0.07 fb to the background. Another
background source is from $2b2cl\slashed{E}_T$, which is approximately
2.5 times as large as that of $4bl\slashed{E}_T$ at the Tevatron. With a $10\%$
mistage rate and after the acceptance cuts, it is reduced to 0.007 fb. 
Finally $3bjl\slashed{E}_T$ and $b3jl\slashed{E}_T$ backgrounds combine to
contribute less than 0.003 fb.

Having tagged three of the four bottom quarks, we identify the
fourth bottom as the hardest untagged jet in the event.
We expect the signal to appear as a peak in the invariant mass
$m(b_1,b_2)$ and $m(b_3,b_4)$ distribution. 
However, 
pairing the four $b$ jets can be complicated due to combinatorics.
We assign the two pairs  by minimizing their mass difference
$m(b_1,b_2)\approx m(b_3,b_4)$
and record both values each with a half weight.
We present the signal versus the background distributions 
of the reconstructed masses 
$m_h$ and $m_a$ in Fig.~\ref{mass4bwTeV} as the
invariant masses of four $b$-jets and of two $b$-jets.
With a simple cut on the $m(4b)$ invariant mass, 
$m(4b) < 160\,\gev$, dictated by
our search for a light Higgs boson with mass smaller than about
$130\,\gev$,
the overall signal to background ratio can be about 10 with 
$C^2=0.50$, $m_a=30\,\gev$ and $m_h=90-130\,\gev$.

To summarize our study at the Tevatron, we claim that the signal
channels of Eqs.~(\ref{wh}) and (\ref{zh}) have 
distinctive kinematical features (see Fig.~\ref{mass4bwTeV}) with negligible 
SM backgrounds and the signal observation is total statistically dominated.
For the $2b 2\tau$ mode, one can reach a cross section of about 0.05$-$0.7 fb
as shown in Fig.~\ref{xsec}, 
while  for the $4b$ mode, we have the cross section in the range of 0.1$-$1 fb
as shown in Fig.~\ref{xsec4b}.
If the $h$ and $a$ masses happen to be related in an optimal 
way (Eq.~(\ref{mamh}))
we can gain an increase in the signal rate by a factor of 
 $\sim 1.8$ and $2.5$ for the $4b$ and $2b2\tau$ channels.
  
\FIGURE[t]{
    \includegraphics[width=0.49\linewidth]{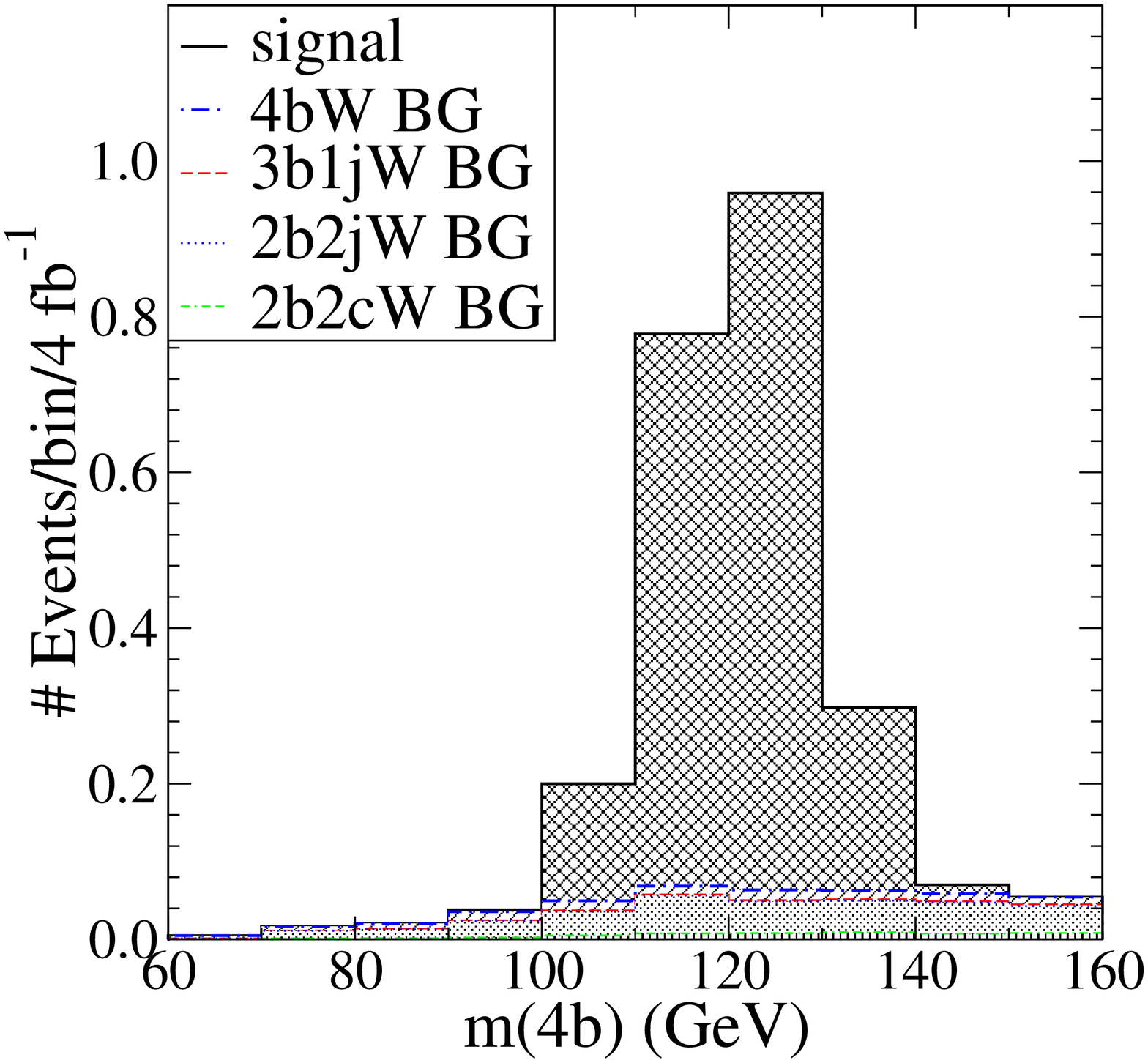}
    \includegraphics[width=0.49\linewidth]{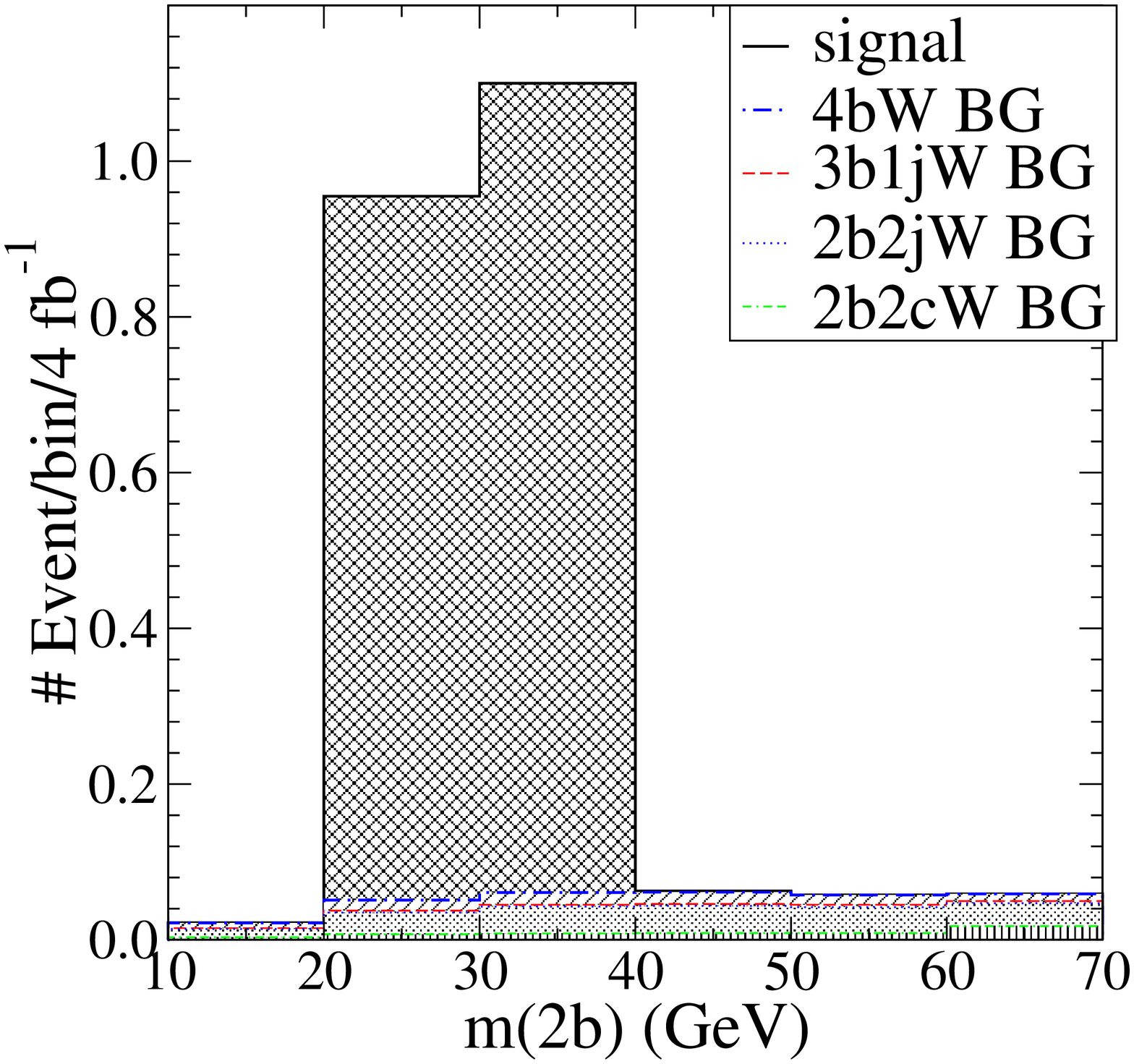}
    \caption {Higgs signal (double-hatched) on top of the sum of the backgrounds 
    at the  Tevatron in the $4b$ decay channel together 
    with a leptonically decaying $W$. 
The invariant mass of four (left) and two (right) $b$-jets are shown.
Values of
$C^2_{4b}=0.50$, $m_h=120\,\gev$and $m_a=30\,\gev$ are understood.
From bottom to top, 
the background histograms indicate the accumulative sum of 
$2b2cW+2b2jW+3b1jW$ and  $2b2cW+2b2jW+3b1jW+4bW$.  }
    \label{mass4bwTeV}
}

\section{$h\to aa$ at the LHC}
\FIGURE[t]{
    \includegraphics[scale=0.34]{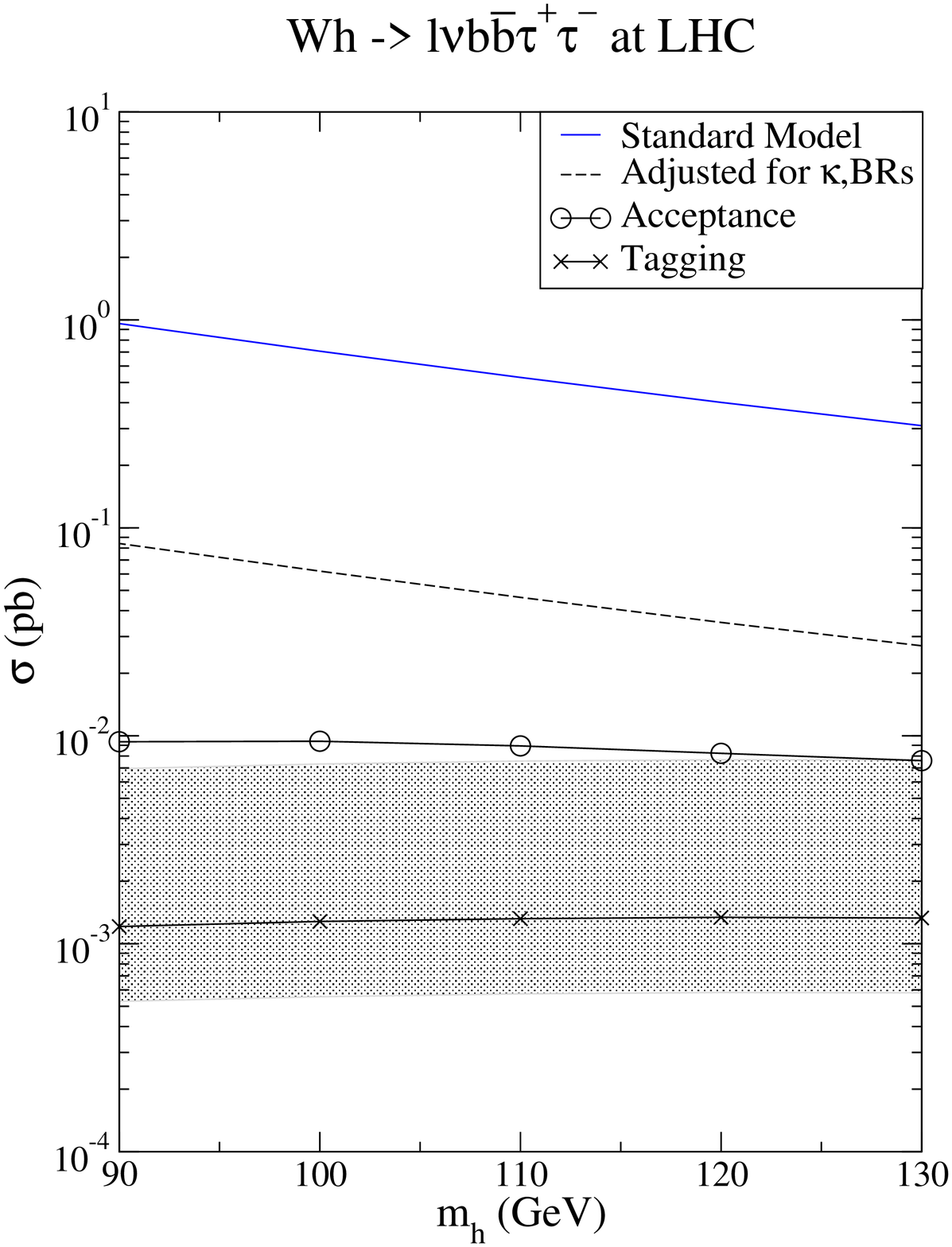}
    \includegraphics[scale=0.34]{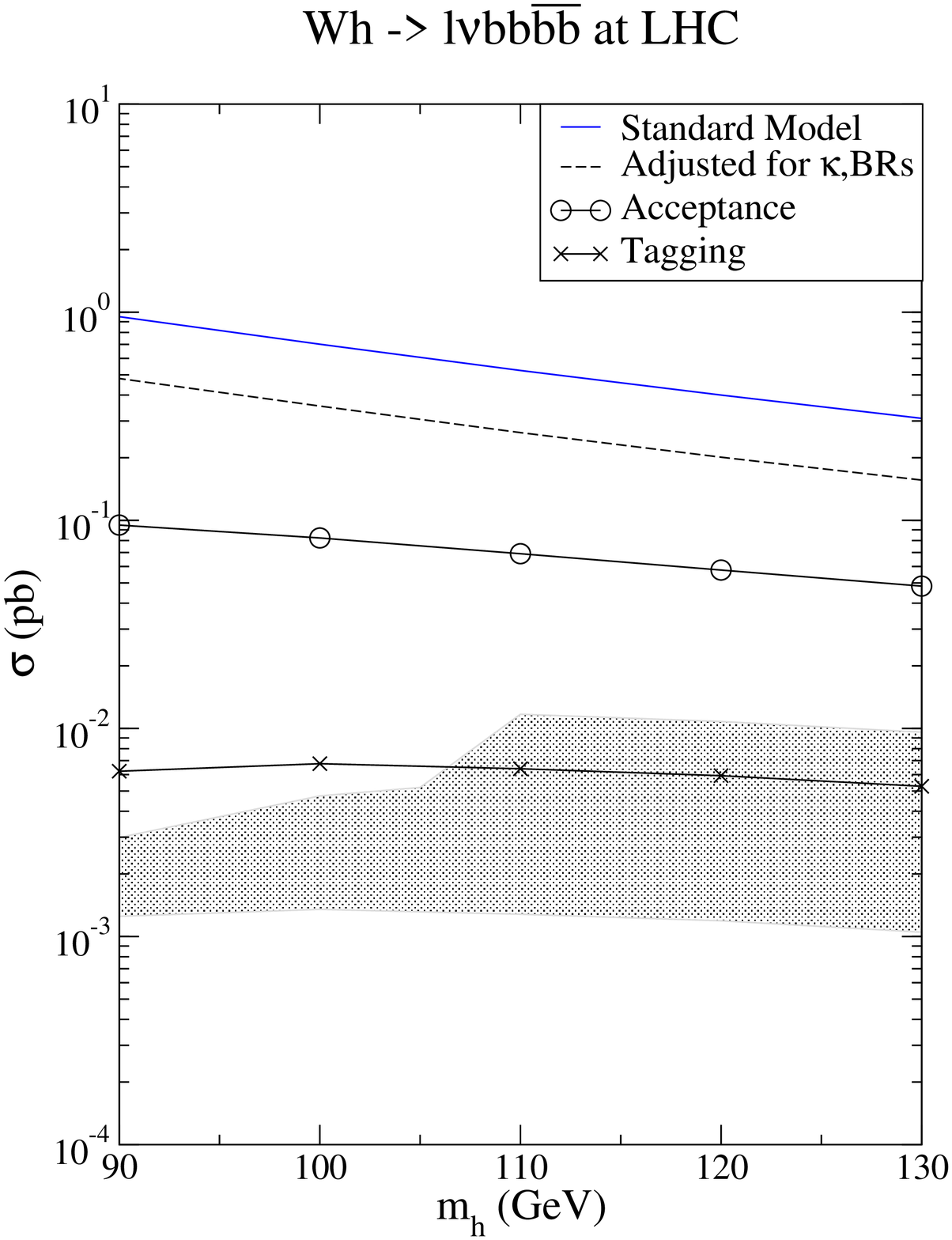}
    \caption{
    Cross sections of Higgs signal at the LHC
        in the $2b2\tau$ (left) and $4b$ (right)
	channels produced by Higgs-strahlung with a leptonically 
	decaying $W$.
The four curves from top to bottom correspond to, respectively, 
the SM cross section (solid line); adjusted for 
$C^2_{2b2\tau}=0.088,\ C^2_{4b}=0.50$ and $m_a=30\,\gev$ (dotted line); 
including the  acceptance cuts Eq.~(\ref{eq:LHCcut}) (line with circles); 
and further including tagging efficiencies Eq.~(\ref{taggingLHC}) (line with crosses).
 The shaded bands  correspond to variations  of the final results (line with crosses)
for values of $C^2$ within the range considered in Table~\ref{param} and 
taking into account the LEP constraints \cite{Schael:2006cr}. }
    \label{xsecwLHC}
}

At the LHC, weak boson-associated Higgs production rate
is about $10-15$ times that at the Tevatron 
in the mass region we are interested in.
With the same $C^2$ factor, 
signal events passing through acceptance also take on this ratio.
The (QCD) background, on the other hand, can be 100 times larger
than at the Tevatron. This requires a substantial jet rejection rate.  
These cross sections are plotted in Fig.~\ref{xsecwLHC}.

Cuts on the triggering leptons and/or missing energy are taken to be
\beq
p_T(l)>20\,\gev, \quad \eta(l)|<2.5, \quad \slashed{E}_T>20\,\gev.
\label{eq:LHCcut}
\eeq
The following cuts and efficiencies for tagging 
are assumed~\cite{unknown:1999fq} 
\bea
\nonumber
\epsilon_b = 50\%  &{\rm for}\ E_T^{jet}>15\,\gev \ &{\rm and} 
\ |\eta_{jet}|<2.0\ , \\
\epsilon_\tau= 40\% &{\rm for}\ E_{vis}>15\,\gev \ &{\rm and} \ |\eta|<2.5\ . 
\label{taggingLHC}
\eea
The jet rejection rate is better than 1/150 for tagging a $b$ or a $\tau$,
except in the $15-30\,\gev$ $p_T$ range where it is taken to be $\sim 1/30$,
as there exists strong tension between tagging efficiencies
and the jet rejection rates, especially near the low $p_T$ range.
Note that the jet rejection rate will only be accurately known after
understanding the detectors with examining the real data.
Again, in our simulations, the energies for a jet and a lepton 
are smeared with  the Gaussian resolutions
\beq
{\Delta E_j \over E_j} = {50\% \over \sqrt{E_j}} \oplus 3\% , \quad
{\Delta E_l \over E_l} = {10\% \over \sqrt{E_l} } \oplus 0.7\% \, .
\eeq
The missing energy is reconstructed accordingly.

\subsection{The $2b2\tau$ Channel}
Similar to the Tevatron case, 
the irreducible background of Eq.~(\ref{irre}) is small after the acceptance
cuts and the tagging requirements, contributing only 0.07~fb.
The reducible background, however,  poses a much more severe problem 
at the LHC. For example, the $2b2jl\slashed{E}_T$ events
are estimated to be around 11~pb, compared to 50~fb at the Tevatron.
Thus for the $2b2\tau$ mode, a jet rejection rate of 1/150 would
give rise to a background of 92~fb, compared to the signal size of about
$1$~fb (or up to $\sim$7~fb when maximizing $C^2$). 
The $4jl\slashed{E}_T$ events also contribute 43~fb to the background
in this channel. 

We carry out the analysis similar to the Tevatron case and arrive at a 
$S/B$ ratio of 0.03, with a total signal size of less than 1~fb 
for $m_h=120\,\gev$, $m_a=30\,\gev$ and $C^2=0.088$.
The small $S/B$ ratio would require precise
control of the systematic errors. It can be further improved by
tagging one more $b$ or $\tau$, at the expense of losing 
up to half of the signal rate. Due to the difficulty of finding
a signal in this channel, we are  led to consider
the more promising channel of $4b$'s,
where, as we did in the Tevatron case, we employ an additional
tagging, while still retaining a higher signal rate.

\subsection{The $4b$ Channel}
With a much higher luminosity than the Tevatron and 
larger cross sections,
LHC could produce 60 (10~$\fbi$) 
to over a thousand (300~$\fbi$) Higgs events
in the $4bl\slashed{E}_T$ decay channel, 
assuming a typical $C^2$ value ($C^2_{4b}=0.50$),
as shown in Fig.~\ref{xsecwLHC}.
The $4b$ channel is thus more optimistic for observing the Higgs,
even though the background still dominates the signal, and
the irreducible $4bl\slashed{E}_T$ 
background becomes non-negligible.
We require tagging three of the $b$ jets, which
would essentially eliminates backgrounds from $4j l\slashed{E}_T$, 
and reduces the $2b2j l\slashed{E}_T$  and $1b3j l\slashed{E}_T$ 
background significantly.
With three tagged $b$-jets, the signal rate is about $5.7$~fb (or up to
10~fb when maximizing $C^2$). The irreducible background  $4bl\slashed{E}_T$ 
is 25 fb. The $3bj l\slashed{E}_T$ background is about 16~fb.
The reducible background from $2b2j l\slashed{E}_T$ events is
about $80$~fb, and $2b2c l\slashed{E}_T$ is about 4 fb with a $10\%$
mistag rate for $c\to b$ \cite{unknown:1999fq}.
The $4j l\slashed{E}_T$ background is no larger than 0.2~fb.

\FIGURE[t]{
    \includegraphics[width=0.49\linewidth]{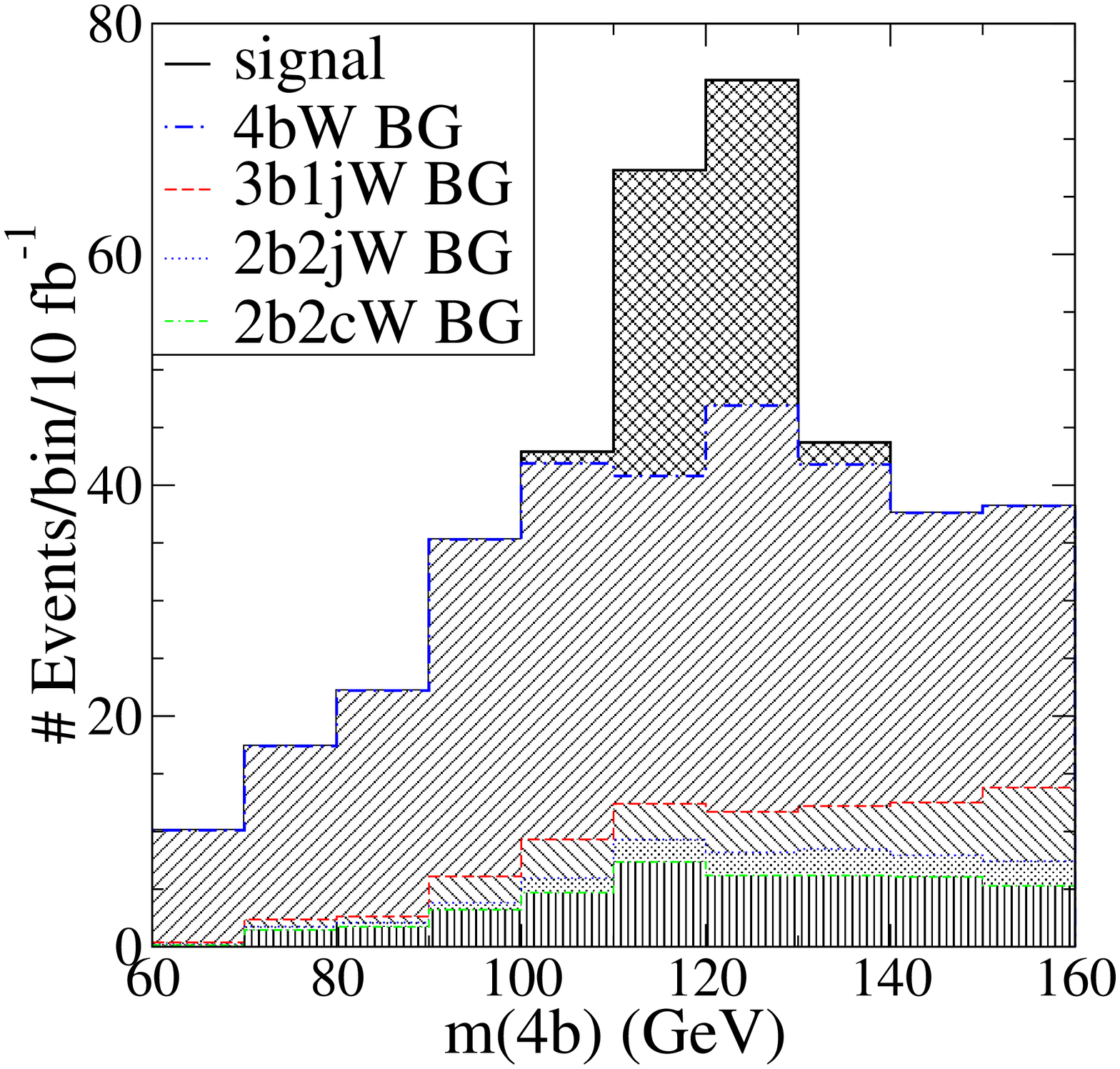}
    \includegraphics[width=0.49\linewidth]{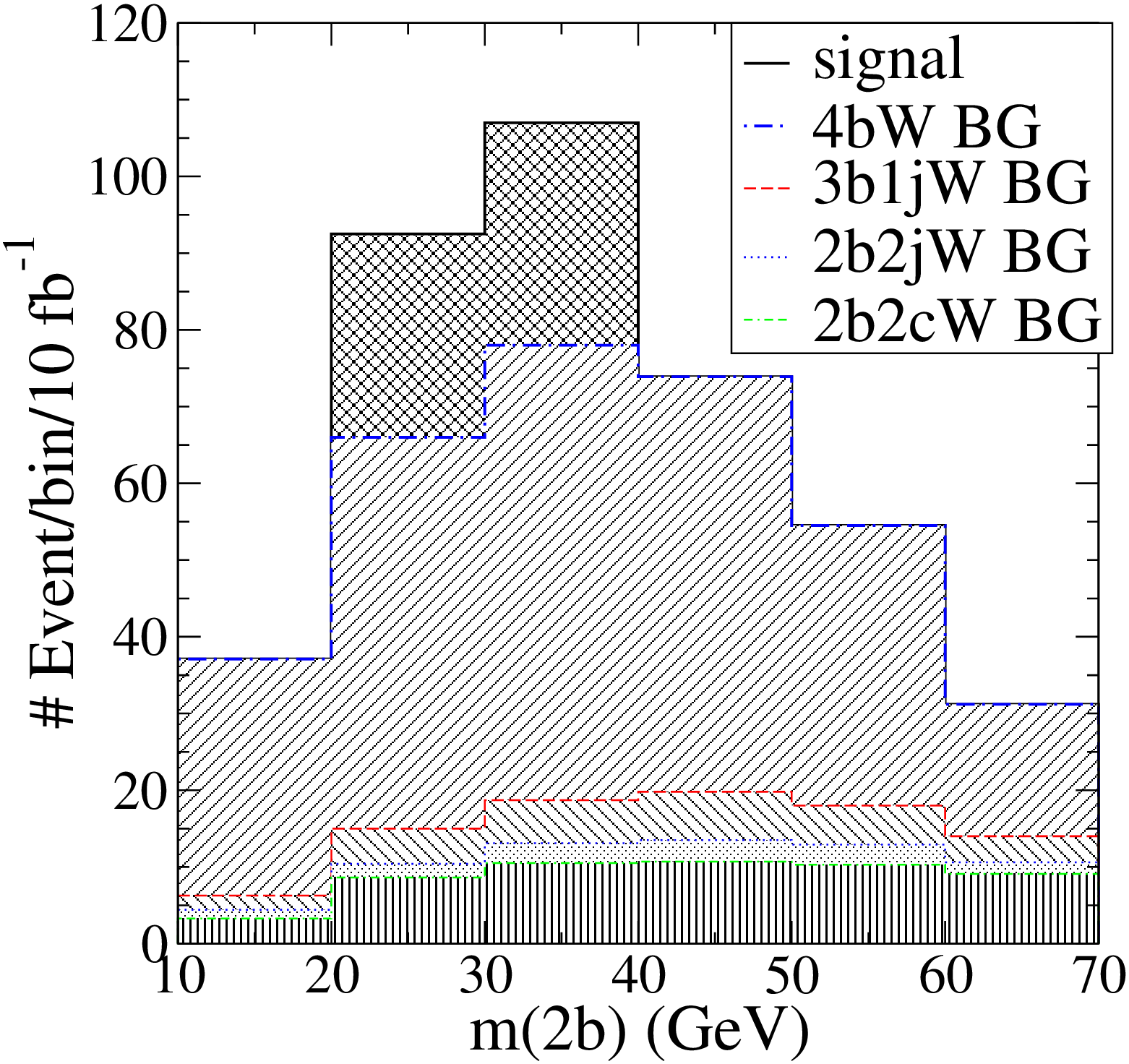}
    \caption {Higgs signal (double-hatched) on top of the sum of the backgrounds at the LHC
 in the $4b$ decay channel together with a leptonically decaying $W$.
The invariant mass of four (left) and two (right) $b$-jets are shown.
Constraints of $60\,\gev < m(4b) < 160\,\gev$ and 
 $10\,\gev < m(2b) < 70\,\gev$ are implemented in both plots.
$C^2_{4b}=0.50$, $m_h=120\,\gev$ and $m_a=30\,\gev$ are understood.
From bottom to top, 
the background histograms indicate the accumulative sum of 
$2b2cW,\ 2b2cW+2b2jW,\  2b2cW+2b2jW+3b1jW$, and 
$2b2cW+2b2jW+3b1jW+4bW$, respectively. }
    \label{mass4bLHCa}
}

We again present the reconstructed mass distribution for the signal 
and backgrounds in two plots in 
Fig.~\ref{mass4bLHCa}.
The left and right plots show the invariant mass distributions of the
$4b$ and $2b$ system, 
where the signal peaks near $m_h=120\,\gev$ and $m_a=30\,\gev$, respectively,
each with a width less than $10\,\gev$ due to detector energy resolution.
Similar to the Tevatron case, we assign the two $bb$ pairs 
by minimizing their mass difference
$m(b_1,b_2)\approx m(b_3,b_4)$
and plot these two masses, each with a half weight.

The dominant $2b2jlE_T$ background comes from $t \bar{t}$
production. For $t \bar{t}$ events, the $2b2j$ system contains
all the decay products of a top-quark. Therefore, these events
may be efficiently rejected with an upper cut on the $m(4b)$
invariant mass lower than the top quark mass,
$m(4b) \simlt 160\,\gev$, which will not affect 
the signal we consider if the Higgs boson mass is in the region 
$m\simlt 130\,\gev$. 
Given our considered range of choices, we implement the following constraints 
in the two distributions:
\bea 
\nonumber
10\,\gev <& m(2b) &< 70\,\gev\,,
\\ \nonumber
60\,\gev <& m(4b) &< 160\,\gev\,.
\eea
While the former affects the $m(4b)$ distribution minimally, the latter
reduces the background in $m(2b)$ distribution by about $40\%$.

Overall, selecting events with these invariant mass constraints,
 the value of $S/B$ is roughly 1/5 for $C^2_{4b}=0.50$. 
Assuming a good understanding of the background,
one can get an estimate of the statistical significance
of the signal. For the rate quoted above we obtain a 
significance, $S/\sqrt{B}$, of over $3.5 \sigma$ for 10 $\fbi$
and over $5\sigma$ for 30 $\fbi$, 
as indicated in Fig.~\ref{mass4bLHCa}. 
If one selects events only in the expected signal region, 
we obtain a $S/B \simeq 0.41$ in the range $100\,\gev < m(4b) < 140\,\gev$
from the $m(4b)$ distribution, 
and a $S/B \simeq 0.40$ in the range $20\,{\gev} < m(2b) < 40\,\gev$
from the $m(2b)$ distribution, 
equivalent to a reduction by about a factor of two
of the luminosity necessary to achieve the same statistical significances.
The challenge is for us to understand the background well enough,
and to control the systematic errors. 

It may be a challenge at the LHC
to retain the high $b$-tagging efficiency at $p_T\sim 15\,\gev$
adopted in the current analysis.
If a $30\,\gev$ cut on the tagged jets is implemented instead,
the signal is reduced to 22\%, while the background drops to about 37\%
of the values given above. 
In such case  a $3\sigma$ ($5\sigma$) signal 
would require an integrated luminosity of around $30\, (80)\, \fbi$. 
Therefore a good understanding of $b$-tagging efficiencies
 at low $p_T$ will be necessary to be able to discover a Higgs in
 the $4b$ channels in the first years of the LHC.

Before closing this section, a remark is in order  for comparing our results
with a recent similar analysis for the $4b$ channel at the LHC \cite{Cheung:2007sva}. 
Their conclusions are somewhat more optimistic, largely due to a significantly higher 
$b$-tagging efficiency assumed ($70\%$). 
They did not consider the QCD  backgrounds of $W2b2c$ and $W2b2j$,
which are sub-leading. On the other hand, 
we neglected the background $t\bar t b\bar b$ considered in  \cite{Cheung:2007sva}
since with the additional energetic $W$
from the top-quark decay $E_W\approx {m^{}_t\over 2} \sqrt{1-M_W^2/m^2_t}$,   
 this background can be efficiently removed
by vetoing the extra jet or charged lepton activities from the $W$ decay.

\section{Summary}
The search for a Higgs boson with couplings to the gauge bosons of
the order of the SM one, and decaying into two lighter CP-odd Higgs bosons 
states may  be performed at hadron colliders for the associate production 
of $Wh,\ Zh$ with $h\to aa\ (2b2\tau\ {\rm or}\ 4b)$.
The cross sections scale proportionally to $C^2$,
a factor determined by the product of the 
relevant branching fractions times the ratio of the 
Higgs production cross section to the SM one. 
Maximal event rates of the two channels are given by different
values of $BR(a\to\tau\tau)$. 
SM-like $af\bar{f}$ couplings tend to give 
small $BR(a\to\tau\tau)$, thus suppressing the $2b2\tau$
channel and enhancing the $4b$ channel.
In models where $BR(a \to \tau\tau)$ is large,
the $2b2\tau$ channel yields an event rate comparable to the $4b$ channel.

We analyzed the $Wh$ channel 
in the mass range $90 \leq m_h\leq\,130\,\gev$ in detail. 
We found that  at the Tevatron
\begin{itemize}
\item
With only basic cuts, the signal size is 0.7 fb for the $2b2\tau$ channel
for $C^2_{2b2\tau} \sim 0.088$ with a negligible  irreducible background,
and $5-10$ fb for the $4b$ channel for $C^2_{4b} \sim 0.50$ with a comparable background. 
With favorable couplings and branching fractions, 
the $C^2$ factor can be as large as 
0.50 for the $2b2\tau$ mode, and 0.90 for the $4b$ mode, and the
signal rate is enhanced proportionally.

\item
 Further cuts and the tagging of $b$ and $\tau$,  
necessary to remove the much larger reducible background, 
worsen the signal event rate to around 0.11~fb for the $2b2\tau$ mode
and 0.5~fb for the $4b$ mode, 
as summarized in Figs.~\ref{xsec} and~\ref{xsec4b}. 
However, the kinematics of the mass reconstruction of $m_a$ and $m_h$ can 
be very distinctive,
as seen in Fig.~\ref{mass4bwTeV}  for the $4b$ mode
with small background and a couple of total events.

\item
We also consider the most favorable relations between $m_h$ and the CP-odd Higgs  
mass $m_a$, which can enhance the signal rate by a factor of 2.5
for the $2b2\tau$ mode leading to a cross section 
as large as 1.6 fb  with $C^2_{2b2\tau} = 0.5$, 
and by a factor of 1.8 for the $4b$ mode leading to a value 1.8~fb 
for $C^2_{4b} = 0.9$.
 
\item
There can be another improvement of $15-30\%$ 
by combining $Wh$ events with the $Zh$ events,
where both $Z\to ll$ and $Z\to \nu\nu$ can be included, 
leading to a possible observation of a few events 
in either $2b2\tau$ or $4b$ channel,
for a Tevatron luminosity of the order of a few $\fbi$.
\end{itemize}
Overall, the signal observation becomes statistically limited. 
Our study has been based on parameters of the CDF detector.
One expects the signal observability to be enhanced accordingly
if results from the D0 detector were combined.

At the LHC, 
the signal rate increases by a factor of 10, 
and the background increases by two orders of magnitude, 
compared to the Tevatron.
We found that
\begin{itemize}
\item
Statistics limitation is no longer a major issue. 
In the $4b$ channel alone the signal rate is 5.7~fb , 
and we can easily obtain a signal significance 
$S/\sqrt{B}$ greater than 3.5
 with an integrated luminosity of 10~$\fbi$, 
and over 10 with 100~$\fbi$. 
\item
Similar to the Tevatron study, with
favorable couplings and branching fractions, 
the signal rate can be enhanced to be as large as 10~fb with $C^2=0.9$,
as seen in Fig.~\ref{xsecwLHC}, and $S/B$ can be improved accordingly. 
\item
The kinematics of the mass reconstruction of $m_a$ and $m_h$ can 
be very distinctive,
as seen in Fig.~\ref{mass4bLHCa}  for the $4b$ mode,
yielding a statistically significant signal.
\item
Favorable $m_h$ and $m_a$ relations, 
combinations of the $Wh$ and $Zh$ signals,
combinations of the $2b2\tau$ channel with the $4b$ channel,
could all improve the signal rate and 
 enhance the potential to the eventual discovery of the Higgs boson.
\end{itemize}
The main challenge would be 
to retain the adequate tagging efficiency of $b$'s and $\tau$'s in the 
low $p_T$ region.

We point out that our background analysis is based on
the leading order partonic calculations in MadEvent. More accurate
estimate of the background distributions would be important
to claim a signal observation. 
More realistic simulations  including the detector effects are needed to
draw more convincing conclusions.

\bigskip
\acknowledgments
TH and G.-Y.H were supported in part by a DOE grant
No. DE-FG02-95ER40896 and in part by the Wisconsin
Alumni Research Foundation.
Work at ANL is 
supported in part by the US DOE, Div.\ of HEP, Contract DE-AC02-06CH11357.
Fermilab is operated by Universities Research Association Inc. under contract 
no. DE-AC02-76CH02000 with the DOE. 

\bibliographystyle{JHEP}
\bibliography{haaj0404}

\end{document}